 \documentclass[final,3p,times,twocolumn]{elsarticle}
\usepackage{amssymb}
\usepackage{amsmath}
\usepackage{url}

\newcommand{\aver}[1]{\left< #1 \right>}

\journal{Physica B}

\begin{document}

\begin{frontmatter}



\title{Influence of pump size on pattern formation 
in exciton-polaritonic Bose-Einstein condensates in the non-Markovian regime}

 \author[label1]{N. V. Kuznetsova}
 \author[label1]{A. D. Alliluev}
 \author[label1]{D. V. Makarov}
 \author[label2]{A. A. Anisich}
 \affiliation[label1]{organization={V.I. Il'ichev Pacific Oceanological Institute of the Far-East Branch of the Russian Academy of Sciences},
             addressline={43 Baltiyskaya Str.},
             city={Vladivostok},
             postcode={690041},
             country={Russia}}
 \affiliation[label2]{organization={Far Eastern Federal University,
 Department of Theoretical Physics},
             addressline={10 Ayax Bay, Russky Island},
             city={Vladivostok},
             postcode={690922},
             country={Russia}}

%\author{N. V. } %% Author name


%% Abstract
\begin{abstract}
Dynamics of exciton-polaritonic condensate under incoherent pumping is studied using the non-Markovian stochastic Gross-Pitaevskii equation with
 the pseudo-differential dispersion term. This term 
corresponds to the lower energy branch of polaritons. 
It is shown that an increasing  of the pumping spot area leads to appearance of various spatial structures whose properties depend
on the duration of the dynamical memory.
In the regime of short memory time, condensate can form an extended state
that spans outside the pumping area. We conclude that onset of such extended
states is related to the specific form of the dispersion term
causing the ``traffic jam'' effect.
The case of long memory time
corresponds to enhanced condensate formation, when increasing of the 
pumping area leads to appearance of angular condensate structures which
partially suppress emission of matter waves from the pumping area.
\end{abstract}



%% Keywords

\end{frontmatter}

\section{Introduction}

Exciton-polaritons are quasiparticles being coupled states of cavity photons and semiconductor excitons.
They have become the object of extensive research due their non-trivial physical properties.
The most prominent one is the extremely low effective mass caused by the impact of the photonic component.
In combination with relatively strong dipole-dipole interaction coming from the excitonic component,
low effective mass aniticipates possibility of high-temperature Bose condensation \cite{Guillet2016}.
Nowadays room-temperature polariton condensation and low-threshold lasing are 
successfully achieved in experiments with organic and perovskite photonic crystals as active media
\cite{Lerario2017,RuiSu2018,RuiSu2020,Zasedatelev2019,Fieramosca2019,Dusel2020,Kolker2024,Wu2024,Song2025,Georgiadis2025},
as well as in colloidal quantum dot microcavities \cite{Dong2025}.
%mainly due to very high values of exciton binding energies.
In addition, exciton-polaritons are considered for implementation of polaritonic lasers \cite{Fabricante2024}, realization of qubits \cite{Ricco2024},
simulation of spin Hamiltonians and topological media \cite{Liew2023}.

Exciton-polaritonic condensate is an intrinsically open system experiencing losses due to photon leakage
and interacting with dense reservoir of excitons. Photon leakage can be compensated by external laser pumping
of the excitonic reservoir, that stimulates relaxation of excitons down to low energies with 
subsequent formation of newborn condensate polaritons.
Stability of condensate density is provided by sufficiently strong coupling to the excitonic reservoir.
Here it is worthwhile to notice that the reservoir has very narrow spectral width in the low-wavenumber range
as the excitonic effective mass
is several orders of magnitude more than the polaritonic one. 
Strong coupling to the reservoir and narrow spectral width
anticipate significance of non-Markovian features
in the condensate-to-reservoir coupling \cite{DeVega}, like the dynamical memory and condensate fluctuations 
being coloured noises with considerable correlation timescales. 
These features can play a remarkable role in temporal and spatial condensate coherence \cite{Brune2025}. 
In turn, macroscopic coherence properties directly govern the spectral linewidth and phase stability of polaritonic lasers, setting fundamental physical limitations for their applications \cite{Fabricante2024}.
Maintaining an ultranarrow linewidth and robust quantum phase is an absolute prerequisite for utilizing polaritonic devices as computational resources in advanced quantum information engineering and continuous-variable protocols \cite{Kavokin2022,Novokreschenov2025}.
Consequently, an accurate non-Markovian description of the reservoir fluctuations and decoherence processes is of critical importance for evaluating the survival lifetime of macroscopic entanglement between spatially separated condensates \cite{Asriyan2026}.
It is reasonable to expect that memory time should depend on environmental temperature.
Some approximations of this dependence were presented in \cite{EL2018,PLA2020,PLA2022,JLTP2024,Bulletin2024,PLA2026}.
A more rigorous treatment was offered in \cite{Quantum}.

The present paper continues the line of the these papers and considers 
metamorphoses of condensate with change of size of the pumping spot.
This problem was recently addressed theoretically and experimentally
in \cite{Utesov}, 
and the qualitative difference between small and large pumping spots
has been underlined. Small pumping spots facilitate outgoing matter waves which become the main mechanism of condensate losses, prevailing over losses due to photonic emission from the microcavity.
The case of large pumping spots corresponds to stronger losses due to photonic emission and the necessary pumping power increases quadratically with spot size.
In the present paper we consider the same problem using the non-Markovian approach, that brings into the problem an additional parameter concerned with the non-Markovian memory. 
In contrast to \cite{EL2018,PLA2020,PLA2022,JLTP2024,Bulletin2024,PLA2026}, we go beyond the parabolic approximation
for the polariton dispersion that is valid only for the range of low wavenumbers.

The paper is organized as follows.
The next section is devoted to brief description of
the non-Markovian 
stochastic Gross-Pitaevskii equation with the modified dispersion term.
The section \ref{sec:Numer} presents
results of numerical simulation.
In Conclusion we summarize and discuss the results obtained.

\section{Theory}
\label{sec:Theory}

In the mean-field approximation for the exciton interaction,
the macroscopic condensate wavefunction of lower polaritons $\psi(\mathbf{r},t)$
obeys the evolution equation of the following form:

\begin{equation}
    i\hbar\frac{\partial\psi}{\partial t} = 
    \hat H_0\psi(\mathbf{r},t) +
\hat D\psi(\mathbf{r},t),
\label{sys0}
\end{equation}
where $\hat H_0$ is the single-polariton Hamiltonian being the unitary part of the right-hand side. Considering only polaritons corresponding to the 
lower energy branch, we can express it as 
\begin{equation}
 \hat H_0 = \hat F^{-1}[E_{\text{LP}}(k)]
    + \alpha_{\text{c}}|\psi(\mathbf{r},t)|^2 + \alpha_{\text{r}}\rho_{\text{r}}(\mathbf{r},t),
    \label{Hamilt}
\end{equation}
where 
$\hat F^{-1}[E_{\text{LP}}(k)]$ is the inverse Fourier transform for dispersion law for the lower polaritons,
$\alpha_{\text{c}}$ is the coupling strength of condensate excitons,
the constant $\alpha_{\text{r}}$ quantifies energy correction due to  coupling of condensate excitons to the excitonic reservoir (the so-called ``blueshift''),
$\rho_{\text{r}}$ is the excitonic reservoir density, and $\hat D$ is the operator describing non-Hermitian interaction of condensate with the environment.
The presence of the operator $\hat F^{-1}[E_{\text{LP}}(k)]$ makes the equation 
(\ref{sys0}) pseudo-differential. 
For the sake of simplicity we consider the case of zero detuning
between excitonic and photonic ground states. In this case
 the dispersion law $E_{\text{LP}}(k)$ is given by the expression
 \begin{equation}
 E_{\text{LP}}(k) = E_0 + \frac{1}{2}\left[
 E_{\text{cav}}(k)
 - \sqrt{E_{\text{cav}}^2(k) + 4\hbar^2\Omega^2}
 \right].
 \label{ELP}
 \end{equation}
 Here $\Omega$ is the Rabi frequency that quantifies energy splitting between polariton spectral branches, and
 $E_{\text{cav}}(k)$ is the dispersion law for cavity photons, that can be readily approximated by the parabolic law,
 \begin{equation}
  E_{\text{cav}}(k) = \frac{\hbar^2k^2}{2m_{\text{cav}}},
 \end{equation}
% %
where $m_{\text{cav}}$ is the effective photon mass in the microcavity.

Non-Hermitian interaction with the environment consists of the photonic and excitonic contributions,
\begin{equation}
 \hat D\psi(\mathbf{r},t) = \hat D_{\text{cav}}\psi(\mathbf{r},t)
 + \hat D_{\text{ex}}\psi(\mathbf{r},t).
\end{equation}
The photonic contribution can be described using the Markov approximation that yields
\begin{equation}
\hat D_{\text{cav}}\psi(\mathbf{r},t) = -i\frac{\hbar\gamma_{\text{cav}}}{2}\psi + \hbar\eta_{\text{cav}}(\mathbf{r},t),
\end{equation}
where $\eta_{\text{cav}}(\mathbf{r},t)$ is spatiotemporal white noise.
Within the truncated Wigner approximation, the autocorrelation function of photonic fluctuations can be written as
\begin{equation}
 \aver{\eta_{\text{cav}}^*(\mathbf{r},t)
 \eta_{\text{cav}}(\mathbf{r},t)
 } = \frac{\gamma_{\text{cav}}}{\Delta x\Delta y}\delta(\mathbf{r} - \mathbf{r'})\delta(t-t'),
\end{equation}
where $\Delta x$ and $\Delta y$ are grid cell sizes.

The excitonic reservoir has relatively narrow energy spectrum and, consequently, relatively long decoherence time,
especially for low temperatures. It means that
interaction with the excitonic reservoir should be essentially non-Markovian.
Here the utilize the model earlier used
in Refs. \cite{PLA2022,JLTP2024}, when the operator $\hat D_{\text{ex}}$ is approximated as
\begin{equation}
 \hat D_{\text{ex}}\psi(\mathbf{r}) \simeq 
 i\frac{\rho_{\text{r}}^2\alpha_{\text{c}}^2 }{\hbar}
 \int\limits_{0}^t dt' \psi(\mathbf{r},t')
 e^{-\gamma_{\text{eff}}(t-t')}\theta(t-t')
 + \eta_{\text{ex}}(\mathbf{r},t),
\end{equation}
where $\theta(t)$ is the Heaviside function,
and the constant $\gamma_{\text{eff}}$ takes into account density of reservoir states which effectively interact with the condensate.
The fluctuation-dissipation theorem yields
the autocorrellation function of excitonic fluctuations,
\begin{equation}
 \aver{\eta^*(\mathbf{r}, t)\eta(\mathbf{r'},t')} = 
 \frac{\rho_{\text{r}}^2\alpha_{\text{c}}^2}{\Delta x\Delta y}
 e^{-\gamma_{\text{eff}}|t-t'|}\delta(\mathbf{r},\mathbf{r'}).
 \label{autocorr}
\end{equation}
Now the equation  (\ref{sys0}) becomes
the non-Markovian stochastic equation of the following form:
\begin{align}
 i\hbar\frac{\partial\psi(\mathbf{r},t)}{\partial t} =
 \hat H_0\psi(\mathbf{r},t) - \frac{i\hbar \gamma_{\text{cav}}}{2}\psi(\mathbf{r},t) \nonumber
 \\
  +  i\frac{\rho_{\text{r}}^2\alpha_{\text{c}}^2 }{\hbar}
 \int\limits_{0}^t dt'\psi(\mathbf{r},t')
 e^{-\gamma_{\text{eff}}(t-t')}\theta(t-t')\\
   + \hbar\eta_{\text{cav}}(\mathbf{r},t) + \hbar\eta_{\text{ex}}(\mathbf{r},t).
 %\hbar\int\limits_{0}^{t}dt' \Sigma^{\text{R}}(t,t')\psi(\mathbf{r},t')
\label{sys1}
 \end{align}
Evolution of the reservoir density is governed by the equation
\begin{align}
&\frac{\partial\rho_{\mathrm{r}}(\mathbf{r},t)}{\partial t} {=} \frac{1}{\hbar}P_{\text{incoh}}(\mathbf{r},t) {-} \gamma_{\text{exR}}\rho_{\mathrm{r}}(\mathbf{r},t) {-}\frac{2}{\hbar}\text{Im}\left[\psi^*(\mathbf{r},t)\eta(\mathbf{r},t)\right] - \nonumber\\
&- \frac{2\rho_{\text{r}}^2\alpha_{\text{c}}^2 }{\hbar^2}\text{Re}\left[
\psi^*(\mathbf{r},t)
\int\,dt' \psi(\mathbf{r},t') e^{-\gamma_{\text{eff}}(t-t')}\theta(t-t')
%\Sigma^{\text{R}}(t,t')\psi(\mathbf{r},t')
\right],
\label{dndt}
\end{align}
where the term $P_{\text{incoh}}(\mathbf{r},t)$ describes the incoherent pumping of the reservoir, $\gamma_{\text{exR}}$ is the decay rate
of the reservoir excitons. The third term on the right-hand side describes polariton exchange between the condensate and the reservoir via the fluctuations.

The exponential form of the memory kernel in (\ref{sys1})
allows to simplify the problem by introducing the auxiliary memory function
\begin{equation}
 \phi(\mathbf{r},t) = \psi_0(\mathbf{r})e^{-\gamma_{\text{eff}}t}
 +\gamma_{\text{eff}}\int\limits_{0}^t\,dt' e^{-\gamma_{\text{eff}}(t-t')}\psi(\mathbf{r},t'),
 \label{fictious}
\end{equation}
where $\psi_0(\mathbf{r}) = \psi(\mathbf{r},t=0)$. 
Evolution of the  auxiliary wave function $\phi$ obeys the equation
 \begin{equation}
 \frac{\partial\phi(\mathbf{r},t)}{\partial t} = \gamma_{\text{eff}}\left[\psi(\mathbf{r},t) - \phi(\mathbf{r},t)\right].
 \label{dmdt}
\end{equation}
Substituting (\ref{fictious}) into (\ref{sys1}) and (\ref{dndt}), we remove time-nonlocality in the right-hand sides, and the equations of condensate and reservoir
evolution take the following form:
\begin{align}
 &i\hbar\frac{\partial\psi(\mathbf{r},t)}{\partial t} =
 \hat H_0\psi(\mathbf{r},t) - \frac{i\hbar \gamma_{\text{cav}}}{2}\psi(\mathbf{r},t) +  \eta(\mathbf{r},t) + \\
 &+i\frac{\alpha_{\text{c}}^2\rho_{\text{r}}^2(\mathbf{r},t)}{\hbar\gamma_{\text{eff}}}
 \left[\phi(\mathbf{r},t) - \psi_0(\mathbf{r})e^{-\gamma_{\text{eff}}t}\right],\label{sys2}\nonumber\\
&\frac{\partial\rho_{\text{r}}(\mathbf{r},t)}{\partial t} = \frac{1}{\hbar}P_{{\text{incoh}}}(\mathbf{r}) - \gamma_{\text{exR}}\rho_{\text{r}}(\mathbf{r},t)\\
&-\frac{2}{\hbar}\text{Im}
\left\{\psi^*(\mathbf{r},t)\eta(\mathbf{r},t)\right\}-\\
-&\frac{2\alpha_{\text{c}}^2\rho_{\text{r}}^2(\mathbf{r},t)}{\hbar^2\gamma_{\text{eff}}}\text{Re}\left\{
\psi^*(\mathbf{r},t)\left[\phi(\mathbf{r},t) - \psi_0(\mathbf{r})e^{-\gamma_{\text{eff}}t}\right]
\right\}.\nonumber
\end{align}
Such procedures are known as Markovian embedding \cite{DeVega,Xiantao,JRLR2022}.

In the present paper we consider the case of a single cw laser beam 
 \begin{equation}
  P_{\text{incoh}}(\mathbf{r}) = \gamma_{\text{exR}}\rho_0   w(\mathbf{r}).
 \end{equation}
Function $w(r)$ describes intensity distribution in the beam. 
It is natural to take it in the Gaussian form,
 \begin{equation}
  w(\mathbf{r}) = \exp\left[-\left(\frac{\mathbf{r} - \mathbf{r_{\text{c}}}}{\sigma_{\text{{r}}}}\right)^{2}
  \right].
 \end{equation}
Our attention is focused on metamorphoses of spatial condensate patterns
with change of the beams size controlled by the parameter $\sigma_{\text{r}}$.

\section{Numerical simulation}
\label{sec:Numer}

%\subsection{Parameter range}
\begin{figure*}[!ht]
\centerline{\includegraphics[width=0.9\textwidth]{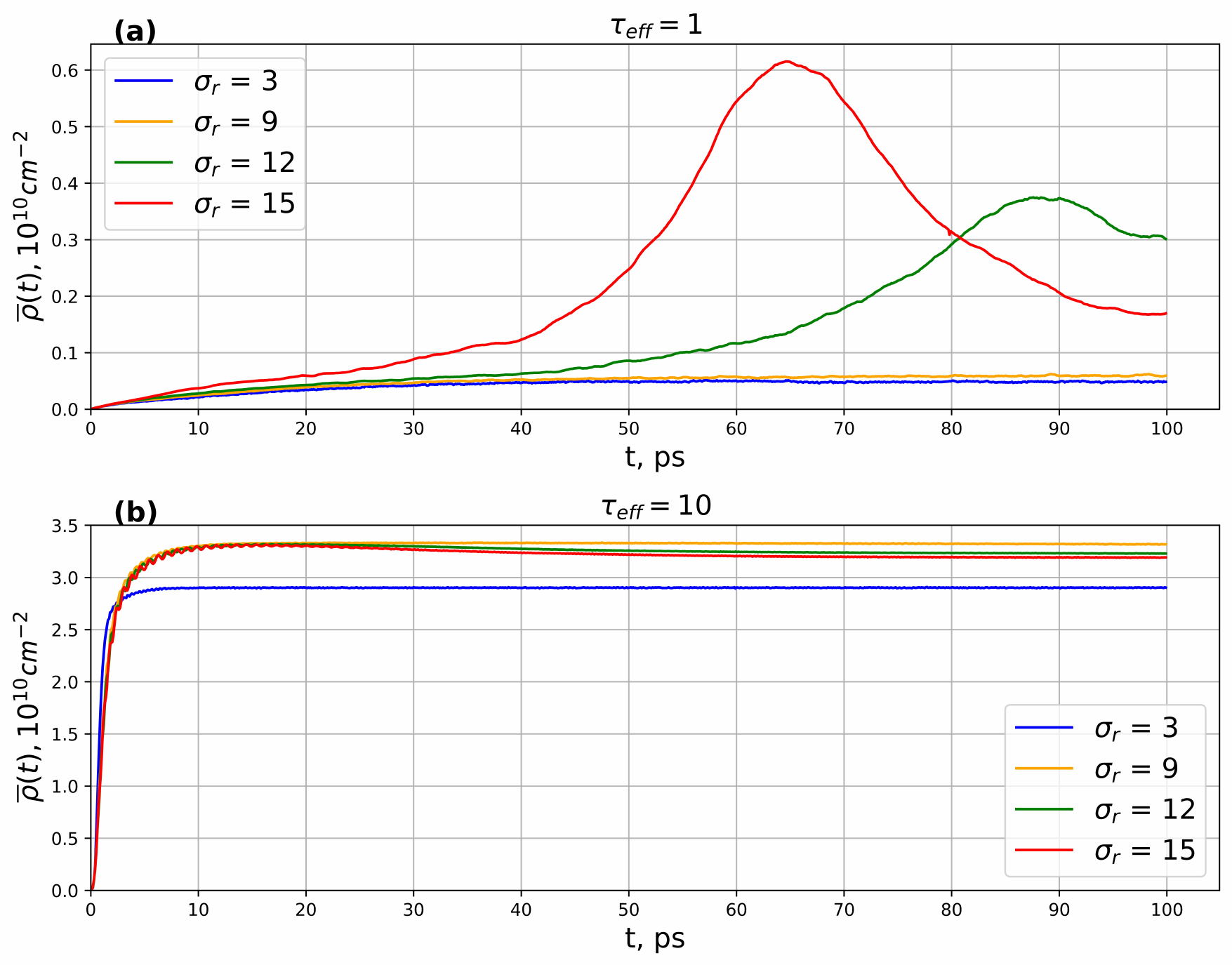}}
\vspace*{8pt}
\caption{Time dependence of averaged condensate density for $\tau_{\text{eff}}=1$~ps (a) and $\tau_{\text{eff}}=10$~ps (b).}
\label{Fig-rho}
\end{figure*}

\begin{figure*}[!ht]
\centerline{\includegraphics[width=0.9\textwidth]{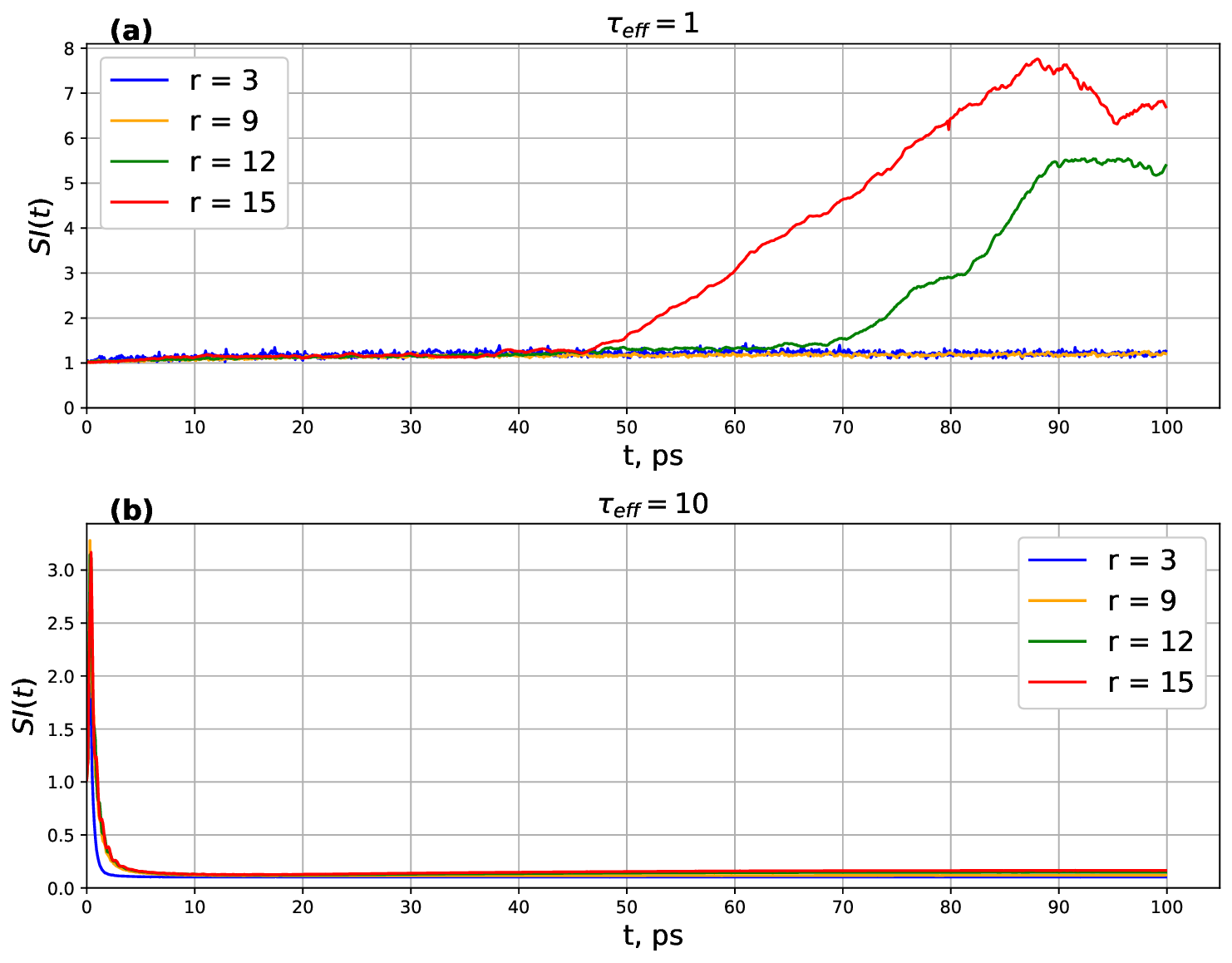}}
\vspace*{8pt}
\caption{Time dependence of the scintillation index quantifying density fluctuations
for $\tau_{\text{eff}}=1$~ps (a) and $\tau_{\text{eff}}=10$~ps (b).}
\label{Fig-SI}
\end{figure*}

The present section is devoted to numerical simulation of condensate dynamics.
In numerical simulation,
the pump spot center is placed at the origin, $\mathbf{r_{\text{c}}} = 0$.
%We consider two values of the parameter $\beta$: $\beta=2$ corresponding to the Gaussian pumping,
%and $\beta=20$ corresponding to the super-Gaussian pumping with nearly cylindrical distribution of intensity.
The value $\rho_0$ represents the maximum reservoir density in the equilibrium state.
We use $\rho_0 = 0.5\times 10^{12}$ $cm^{-2}$ for numerical simulation.

%The parameter $\sigma_{\text{r}}=20$ $\mu m$ sets the half-width of the pump spot.
The main goal of the present paper is to study coherence properties of the condensate for various sizes of the pumping spot.
As it was shown in \cite{PLA2022,JLTP2024} condensate coherence 
is controlled by the memory timescale
$\tau_{\text{eff}}=1/\gamma_{\text{eff}}$.
We consider the values of $\tau_{\text{eff}}$ ranging from 1 ps (short memory)
to 10 ps (relatively long memory).
The values of $\gamma_{\text{cav}}$,  and $\gamma_{\text{exR}}$ and are determined by the corresponding lifetimes:
$\tau_{\text{cav}}=1/\gamma_{\text{cav}}=3.8$~ps, 
  $\tau_{\text{exR}}=1/\gamma_{\text{exR}}=10$~ps.
The interexciton interaction constant $\alpha_{\text{c}}$ is set to $6\cdot 10^{-14}$~eV$\cdot$cm$^{2}$, the size of the grid cell $\Delta x=\Delta y=0.5$ $\mu m$.
We consider the case of zero initial conditions
\begin{equation}
 \psi(t=0) = \phi(t=0) = \rho_{\text{r}}(t=0) = 0.
\end{equation}
%

%\subsection{Condensate density}

\begin{figure*}[!ht]
\centerline{\includegraphics[width=0.98\textwidth]{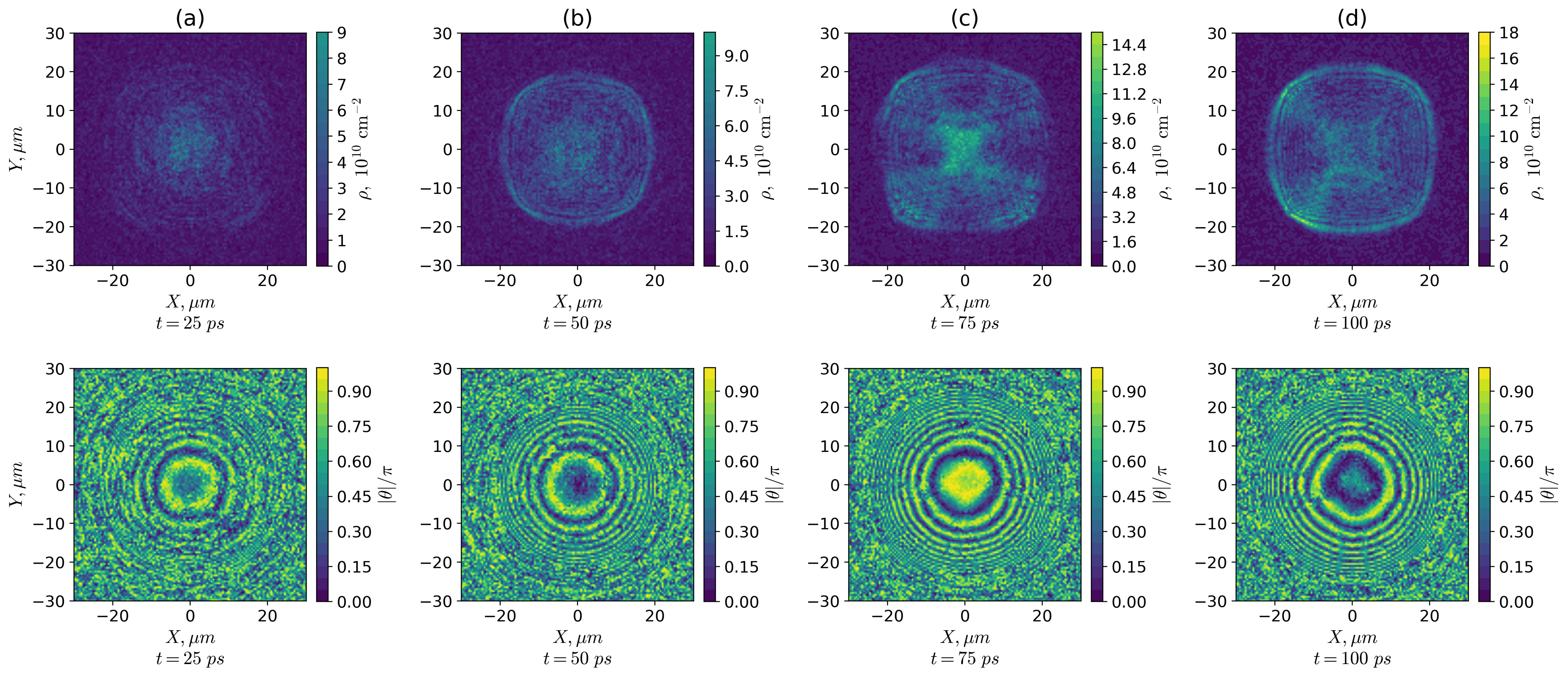}}
\vspace*{8pt}
\caption{Snapshots of density and phase distributions at various time instants
for a typical individual realization of spatiotemporal fluctuation fields.
The case of $\sigma_r=15$~$\mu$m and $\tau_{\text{eff}}=1$ ps.
}
\label{Fig-Snapshots01}
\end{figure*}

\begin{figure*}[!ht]
\centerline{\includegraphics[width=0.98\textwidth]{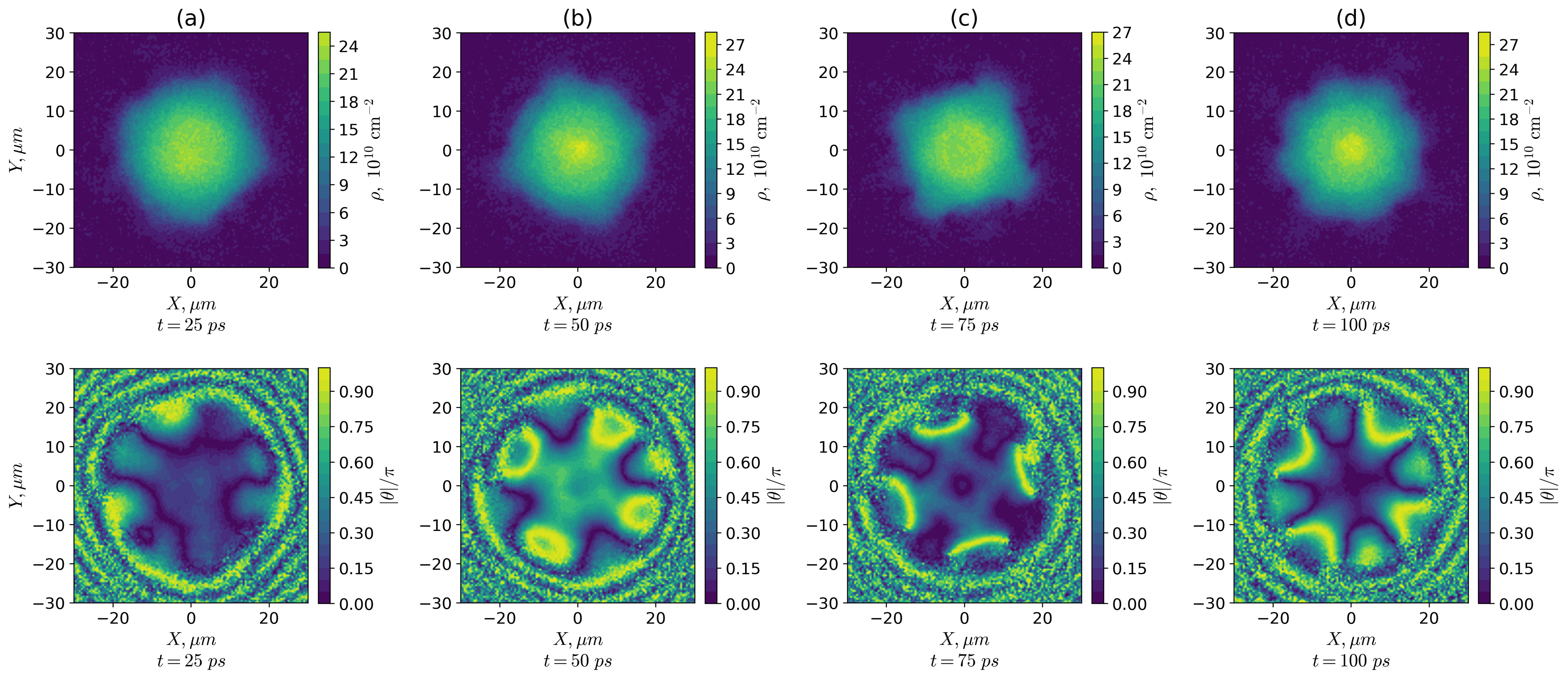}}
\vspace*{8pt}
\caption{Snapshots of density and phase distributions at various time instants
for a typical individual realization of spatiotemporal fluctuation fields.
The case of $\sigma_r=15$~$\mu$m and $\tau_{\text{eff}}=10$ ps.
}
\label{Fig-Snapshots10}
\end{figure*}

Condensate density is one of the main indicators of its state. Indeed, high enough density ensures
suppression of inevitable quantum fluctuations and stability of condensate phase.
In the present paper we carry out weighted spatial average of density,

\begin{equation}
 \bar\rho =\int d\mathbf{r}\, f(\mathbf{r})\aver{|\psi(\mathbf{r})|^2}\,
\end{equation}
where angular brackets denote the ensemble average, and the weighting function $f(\mathbf{r})$ is determined
by the pumping spot form,
\begin{equation}
    f(\mathbf{r}) = \frac{w(\mathbf{r})}{\int w(\mathbf{r})d\mathbf{r}}.
    %     \Biggl\{\Biggr.\begin{aligned}
%     1,\quad |\mathbf{r}-\mathbf{r}_{\text{c}}|\le \sigma, \\
%     0, \quad |\mathbf{r}-\mathbf{r}_{\text{c}}| > \sigma.
%     \end{aligned}
\end{equation}
Fluctuations of condensate density can be quantified by means of the scintillation index (SI) defined as
\begin{equation}
 \text{SI} = \frac{\aver{\bar\rho^2}}{\aver{\bar\rho}^2} - 1.
\end{equation}
One has $\text{SI}=0$ for the absence of fluctuations, and $\text{SI}=1$ for the regime of statistically saturated fluctuations.
The range $\text{SI}>1$ corresponds to regime of strong density fluctuations.

Figures \ref{Fig-rho} demonstrates temporal variations of averaged density. 
In the case of short memory time formation of considerable condensate density is observed only within a limited time interval
(see Fig.~\ref{Fig-rho}(a)) and only for relatively large pumping spots. According to data presented in Fig.~\ref{Fig-SI}(a),
these temporal onsets of condensate are followed by fast amplification of density fluctuations, that indicates
on stochastic behavior of condensate inside the pumping spot. 
Figure \ref{Fig-Snapshots01} represents a typical example of density and phase distributions
for $\sigma_r=15$~$\mu$m and $\tau_{\text{eff}}=1$ ps, and one can see that
the condensate occupies wide domain with sharp boundaries and constructive 
interference pattern within. The domain area significantly exceeds the area of the pumping spot. The interference pattern is somewhat similar to
``sunflower ripples'' observed in \cite{Sunflower}.
Radial matter waves propagating between the domain center and
the domain boundary lead to transport of density from the center to the domain periphery, and this transport is accompanied by strong fluctuations of
$\bar\rho$. 
It is worth noticing that wavenumber of these waves increases with 
increasing the distance from the pumping spot center,
evidently due to action of the repulsive forces associated
with exciton-exciton interaction. According to the form
of the dispersion term (\ref{ELP}),
increasing of wavenumber leads to decreasing of wave velocity.
Thus, there occurs the effect of ``traffic jam'' corresponding to
accumulation of condensate density at some outlying region that becomes
the boundary of the resulting extended state.

Elongation of the memory time
$\tau_{\text{eff}}$ from 1 to 10 ps leads to drastic increase of condensate density. The presence of strong dynamical memory facilitates onset of 
spatial phase coherence that corresponds to favorable conditions for 
pumping from the reservoir.
Therefore the condensate rapidly reaches an equilibrium state
with sufficiently high density and very weak fluctuations 
(see Figs.~\ref{Fig-rho}(b) and \ref{Fig-SI}(b)).
Notably, the equilibrium density for the smallest pumping spot 
with $\sigma_r=3$ $\mu$m is lower
due to increased losses associated with enhanced emission 
of outgoing ballistic matter waves \cite{Utesov}.
As the pumping area increases, width of condensate spatial spectrum
decreases that diminishes excitation of ballistic states.
On the other hand, it facilitates long-wavelength excitations
that results in onset of angular patterns like four-fold phase islands
presented
in Fig.~\ref{Fig-Snapshots10}. It should noticed that these patterns
are not associated with vortices: there are no corresponding phase singularities. Each of the phase islands experiences density pulsations
which cause specific deformations of the condensate spot.
Angular locations of phase islands depend of realization
of a fluctuation field and are random.

\begin{figure*}[!ht]
\centerline{\includegraphics[width=0.98\textwidth]{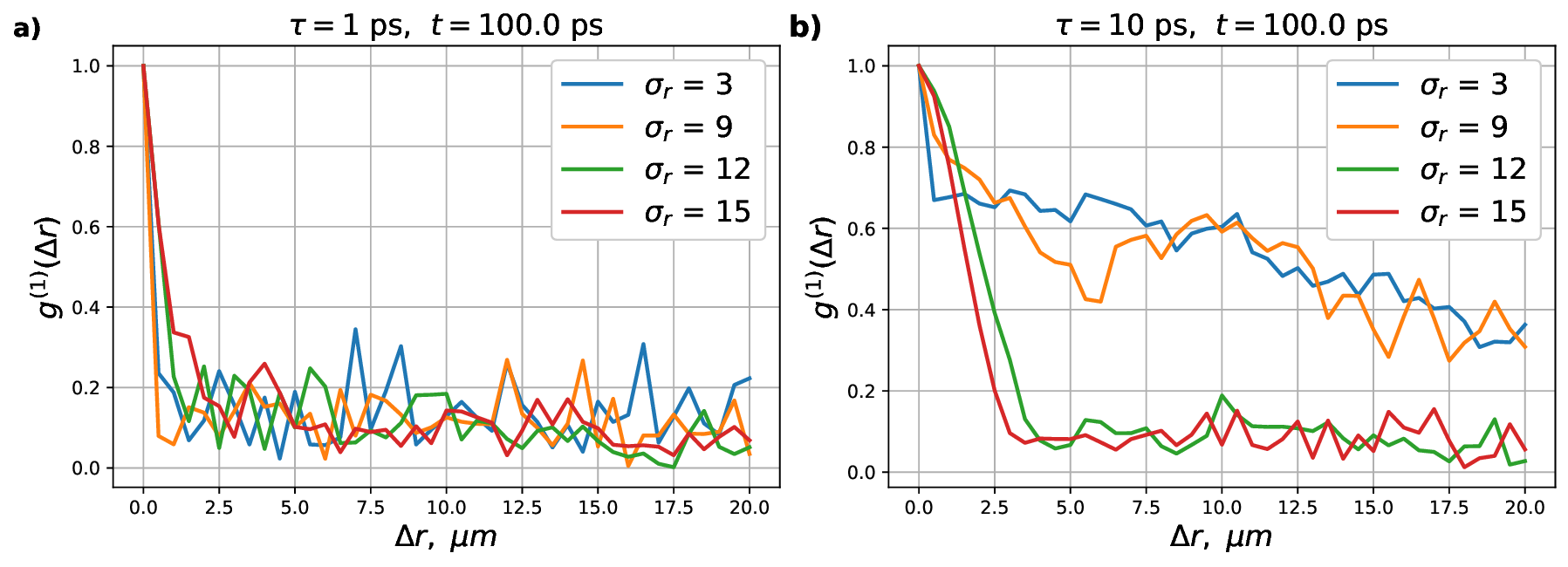}}
\vspace*{8pt}
\caption{The first-order coherence function $g^{(1)}(\Delta r)$
for various values of $\sigma_r$ and $\tau_{\text{eff}}$.
}
\label{Fig-G1}
\end{figure*}

Spatial phase coherence of condensate can be quantified by means of 
the first-order coherence function $g^{(1)}(\Delta\mathbf{r},t)$ that can defined as
 \cite{RuiSu2020,JLTP2024,Fontaine}
\begin{equation}
    g^{(1)}(\Delta\mathbf{r},t) =
    \frac{|\aver{\psi^*(\mathbf{r_0+\Delta r/2},t)\psi(\mathbf{r_0-\Delta r/2},t)}|}
    {\aver{\sqrt{\rho_{\text{c}}(\mathbf{r_0+\Delta r/2},t)\rho_{\text{c}}(\mathbf{r_0-\Delta r/2},t)}}},
\label{eq:g1}
\end{equation}
with $\rho_{\text{c}}(\mathbf{r},t) = |\psi(\mathbf{r},t)|^2$ being the condensate density.
We have fixed $\mathbf{r_0}=\mathbf{0}$ and average over the two lines: the vertical ($x=0$) and horizontal ($y=0$).

Fig,~\ref{Fig-G1} represents the curves $g^{(1)}(\Delta\mathbf{r},t)$.
In the case of $\tau_{\text{eff}}=1$~ps the first-order coherence 
function rapidly drops down due to strong impact of excitonic and photonic fluctuations.
A qualitatively different picture is observed for $\tau_{\text{eff}}=10$~ps,
when condensates created by the small pumping spots exhibit
very slow decay of spatial correlations due to extensive coherent emission of outgoing ballistic matter waves. 
Increasing of the pumping spot area strongly
diminishes the emission. It turns out that the onset of angular island-like patterns leads to drop-down of spatial phase correlations.
Here it should be taken into account that angular locations of these patterns
are random.
 Consequently, they
give incoherent contribution into calculation of 
$g^{(1)}(\Delta\mathbf{r})$. It results in accelerated decay of
$g^{(1)}(\Delta\mathbf{r})$.

% %
%     \begin{equation}
%         \Psi_{\text{eff}}(t) = 
% % \frac{1}{\gamma_{\text{exR}}\rho_0}\int\int 
%         \frac{\int d\mathbf{r} w(\mathbf{r})\psi(\mathbf{r},t)}{\int d\mathbf{r} w(\mathbf{r})}.
%     \end{equation}
% %
% To obtain the spectrum of the signal, we use the Gabor transform, which is a variation of the Fourier window transform with the Gaussian window:
% %
%     \begin{equation*}
%         g_{W}(\nu,t) = \int_{-\infty}^{\infty}dt^{\prime} \Psi_{\text{eff}}(t^{\prime})\chi_{\text{W}}(\nu,t^{\prime}-t),
%     \end{equation*}
% %    
% where $\chi_{\text{W}}(\nu,t) = R_{\text{W}}(t)\exp{(2\pi i\nu t)}$ and
% %
%     \begin{equation}
%         R_{\text{W}}(t) = \frac{1}{\sqrt{W\sqrt{2\pi}}}\exp{\left(-\frac{t^2}{4W^2}\right)}.
%     \end{equation}
% %    
% 
% 
% The window is centred at $t = t^{\prime}$, and $W$ sets its size. By moving the centre of the window along the time axis, one may trace the time dependence of the frequency spectrum of $\Psi_{\text{eff}}(t)$. 

\section{Conclusion}
\label{Concl}

In the present paper we study metamorphoses of exciton-polaritonic 
Bose-Einstein condensates that occur with increasing of the 
pumping spot area. We utilize the non-Markovian 
stochastic Gross-Pitaevskii equation in 
the pseudo-integral form that takes into account actual dispersion relation
for lower polaritons.
It is found that increasing of pumping spot area significantly
enriches spatial condensate structure.
In the short-memory regime that corresponds to relatively high
temperatures, there occur extended condensate states with 
sharp boundaries. We anticipate that these boundaries result from the 
``traffic jam'' effect caused by decreasing of velocity
of condensate waves outflowing from the pumping spot.
These extended states have lifetimes of few tens ps
and experience strong
density fluctuations. 

In the long-memory regime corresponding
to low temperatures, condensate density and phase coherence are significantly higher. Small-size pumping spot creates condensate with 
lesser density due to enhanced emission of the outflowing ballistic waves.
Increasing of the pumping spot size results in onset of angular 
island-like phase patterns that cause boundary deformations of the condensate spot and reduce spatial phase coherence.

The main question that still has to be answered is possibility of experimental observation of the aforementioned structures. 
It is natural to assume that
the memory time is linked to temperature, 
but accurate form of the link is still lacking.
In our earlier studies, this link was derived under the assumption 
of Boltzmann distribution of reservoir excitons, but it seems 
to be very rough assumption.
A correct analytical form of the memory dependence on temperature and its influence on condensate dynamics
shall be the object of our further research.

\section*{Acknowledgements}
The work is supported by
the project No. 124022100072-5 at the Pacific Oceanological Institute of FEB RAS, and by the Foundation of the Advancement of Theoretical Physics and Mathematics ``Basis''.

% \section*{ORCID}
% You are encouraged to include in your user information the ORCID (\url{https://orcid.org/}) or register for one if you don't have it. This ID will help to identify you in the researcher community and make it easier to keep track of all your publications.
% Please provide a valid ORCID here, e.g.,

% \noindent Josiah Carberry - \url{https://orcid.org/0000-0002-1825-0097}
% 
% \noindent Rajesh Babu - \url{https://orcid.org/0009-0006-0415-6880}

% \noindent
% Typeset references in 9pt Times Roman.
\vspace*{3pt}

%N.V. Kuznetsova: Investigation, Methodology, Formal analysis, Software, Visualization, Writing - Original Draft, Writing - Review \& Editing.
%A.D. Alliluev: Investigation, Formal Analysis, Methodology, Writing - Original Draft, Writing - Review \& Editing.
%D.V. Makarov: Conceptualization, Methodology, Software, Writing - Original Draft, Writing - Review \& Editing.
%A.I. Anisich: Investigation, Formal Analysis, Methodology, Writing - Review \& Editing.


\begin{thebibliography}{50}

%
\bibitem{Guillet2016} 
T.~Guillet and C.~Brimont, Polariton condensates at room temperature,
Comp. Rend. Phys. 17 (2016) 946.
{\url{https://doi.org/10.1016/j.crhy.2016.07.002}}.
%
\bibitem{Lerario2017}
G.~Lerario, A.~Fieramosca, F.~Barachati, D.~Ballarini,
K.S.~Daskalakis, L.~Dominci, M.~De Giorgi, S.~A.~Maier, 
G.~Gigli, S.~K\'ena-Cohen, and D.~Sanvitto,
Room-temperature superfluidity in a polariton condensate,
Nature Physics 13 (2017) 837. 
\url{https://doi.org/10.1038/nphys4147}.
%
\bibitem{RuiSu2018}
R.~Su, J. Wang, J. Zhao, J. Xing, W. Zhao, C. Diederichs, T. C. H.~Liew,
Q.~Xiong,
Room temperature long-range coherent exciton polariton condensate flow in lead halide perovskites,
Science Advances 4 (2018) eaau0244. 
\url{https://doi.org/10.1126/sciadv.aau0244}.
%
\bibitem{Zasedatelev2019}
A. V.~Zasedatelev, A. V. Baranikov, D. Urbonas,
F. Scafirimuto, U. Scherf, T. St\"oferle, R. F. Mahrt,
and P. G. Lagoudakis, 
A room-temperature organic polariton transistor,
Nature Photonics 13 (2019) 378. 
\url{https://doi.org/10.1038/s41566-019-0392-8}.
%
\bibitem{Fieramosca2019}
A.~Fieramosca, L.~Polimeno, V.~Ardizzone, L. De Marco, M.~Pugliese,
V.~Maiorano, M.~de Giorgi, G.~Gigli, D.~Gerace, D.~Ballarini, D.~Sanvitto,
Two-dimensional hybrid perovskites sustaining strong polariton interactions at room temperature,
Science Advances 5 (2019) aav9967 (2019).
\url{https://doi.org/10.1126/sciadv.aav9967}.
%
\bibitem{Dusel2020}
M. Dusel, S. Betzold, O. A. Egorov, S. Klembt, J. Ohmer, U. Fischer, 
S. H\"0fling, and C. Schneider,
Room temperature organic exciton–polariton condensate in a lattice,
Nature Communications 11 (2020) 2863. 
\url{https://doi.org/10.1038/s41467-020-16656-0}.
%
\bibitem{RuiSu2020}
R.~Su, S. Ghosh, J. Wang, S. Liu, C. Diederichs, T. C. H. Liew, and Q. Xiong,
Observation of exciton polariton condensation in a perovskite lattice at room temperature,
Nature Physics 16 (2020) 301. 
\url{https://doi.org/10.1038/s41567-019-0764-5}.
%
\bibitem{Kolker2024}
M. D. Kolker, I. I. Krasionov, A. D. Putintsev, E. D. Grayfer, T. Cookson, D. Tatarinov, A. P. Pushkarev, D. A. Sannikov, P. G. Lagoudakis
Room temperature broadband polariton lasing from a CsPbBr3 perovskite plate,
Adv. Optic. Mater.13 (2024) 2402543.
\url{https://doi.org/10.1002/adom.202402543}.
%
\bibitem{Wu2024}
X. Wu, S. Zhang, J. Song, X. Deng, W. Du, X. Zeng, Y. Zhang, Z. Zhang, Y. Chen, Y. Wang, C. Jiang, Y. Zhong, B. Wu, Z. Zhu, Y. Liang, Q. Zhang, Q. Xiong, and X. Liu,
Exciton polariton condensation from bound states in the continuum at room temperature,
Nature Commun. 15 (2024) 3345. 
\url{https://doi.org/10.1038/s41467-024-47669-8}.
%
\bibitem{Song2025}
J. Song, S. Ghosh, X. Deng, C. Li, Q. Shang, X. Liu, Y. Wang, 
X. Gao, W. Yang, X. Wang,
Room-temperature continuous-wave pumped exciton polariton condensation in a perovskite microcavity,
Sci. Adv. 11 (2025) eadr1652. 
\url{https://doi.org/10.1126/sciadv.adr1652}.
%
\bibitem{Georgiadis2025}
I, Georgakilas, D. Tiede, D. Urbonas, R. Mirek, C. Bujalance, L. Cali\'0, 
V. Oddi, R. Tao, D. N. Dirin, G. Rain\'0, S. C. Boehme, J. F. Galisteo-L\'opez, R. F. Mahrt, M. V. Kovalenko, H. Miguez, and T St\"oferle,
Room-temperature cavity exciton-polariton condensation in perovskite quantum dots,
Nature Commun. 16 (2025) 5228 (2025).
\url{https://doi.org/10.1038/s41467-025-60553-3}.
%
\bibitem{Dong2025}
J. Dong, Y. Wu, R. Wang, L. Wang, J. Wang, Y. Zhang, Y. Wang, X. Wang,
S. Shen, and H. Zhu,
Low-threshold colloidal quantum dot polariton lasing via a strong coupling microcavity at room temperature,
Nanoscale 17 (2025) 10187. 
\url{https://doi.org/10.1039/d4nr05185h}.
%
\bibitem{Fabricante2024}
B. R. Fabricante, M. Kr\'ol, M. Wurdack, M. Pieczarka, M. Steger, 
D. W. Snoke, K. West, L. N. Pfeiffer, A. G. Truscott, E. A. Ostrovskaya, 
and E. Estrecho,
Narrow-linewidth exciton-polariton laser,
Optica 11 (2024) 838. 
\url{https://doi.org/10.1364/OPTICA.525961}.
%
\bibitem{Ricco2024}
L.S.~Ricco, I.A.~Shelykh, and A.~Kavokin, 
Qubit gate operations in elliptically trapped polariton condensates,
Sci. Rep. 14 (2024) 4211.
\url{https://doi.org/10.1038/s41598-024-54543-6}.
%
\bibitem{Liew2023}
T.C.H.~Liew, 
The future of quantum in polariton systems: opinion,
Opt. Mater. Express 13 (2023) 2085. 
\url{https://doi.org/10.1364/OME.492503}.
%
\bibitem{DeVega}
I.~De Vega, D.~Alonso, 
Dynamics of non-Markovian open quantum systems,
Rev.~Mod.~Phys. 89 (2017) 015001.
\url{https://doi.org/10.1103/RevModPhys.89.015001}.
%
\bibitem{Brune2025}
Y. Brune, E. Rozas, K. West, K. Baldwin, L. N. Pfeiffer, J. Beaumariage, H. Alnatah, D. W. Snoke, and M. Aßmann, 
Quantum coherence of a long-lifetime exciton-polariton condensate,
Commun. Mater. 6 (2025) 1. 
\url{https://doi.org/10.1038/s43246-025-00848-6}.
%
\bibitem{Kavokin2022}
A. Kavokin, T. C. H. Liew, C. Schneider, P. G. Lagoudakis, S. Klembt, and 
S. Hoefling,
Polariton condensates for classical and quantum computing, 
Nat. Rev. Phys. 4 (2022) 435.
\url{https://doi.org/10.1038/s42254-022-00447-1}.
%
\bibitem{Novokreschenov2025}
D.~Novokreschenov, A.~Kudlis, and A.V.~Kavokin, 
Classical and single photon memory devices based on polariton lasers, 
arXiv preprint (2025), 
\url{https://arxiv.org/abs/2509.20569}.
%
\bibitem{Asriyan2026}
N.A.~Asriyan, A.A.~Elistratov, and A.V.~Kavokin, 
Generating entangled polaritonic condensates by pumping with entangled pairs of photons,
arXiv preprint (2026), \url{https://arxiv.org/abs/2602.22778}.
%
\bibitem{EL2018}
A.A.~Elistratov, Yu.E.~Lozovik, 
Polariton Bose condensate in an open system: Ab initio approach,
Phys. Rev. B 97 (2018) 014525.
\url{https://doi.org/10.1038/s41467-025-60553-3}.
%
\bibitem{PLA2020}
D.V.~Makarov, A.A.~Elistratov, Yu.E.~Lozovik, 
Non-Markovian effects in dynamics of exciton-polariton Bose condensates,
Phys. Lett. A 384 (2020) 126942.
\url{https://doi.org/10.1016/j.physleta.2020.126942}.
%
\bibitem{PLA2022}
A.D.~Alliluev, D.V.~Makarov, N.A.~Asriyan,
A.A.~Elistratov, Yu.E.~Lozovik, 
Formation of exciton-polaritonic BEC in the non-Markovian regime,
Phys. Lett. A 453 (2022) 128492.
\url{https://doi.org/10.1016/j.physleta.2022.128492}.
%
\bibitem{JLTP2024}
A.D.~Alliluev, D.V.~Makarov, N.A.~Asriyan,
A.A.~Elistratov, Yu.E.~Lozovik, 
Non-Markovian stochastic Gross–Pitaevskii equation for the exciton–polariton Bose–Einstein condensate,
J. Low Temp. Phys. 214 (2024) 331.
\url{https://doi.org/10.1007/s10909-023-03027-4}.
%
\bibitem{Bulletin2024}
N.~V.~Kuznetsova, D.~V.~Makarov, N.~A.~Asriyan,
A.~A.~Elistratov, Yu.~E.~Lozovik, 
Spatial coherence of exciton–polariton Bose–Einstein condensates,
Bull. Russ. Acad. Phys. 88 (2024) 847, 
\url{https://doi.org/10.1134/S106287382470672X}.
%
\bibitem{PLA2026}
N.V.~Kuznetsova, D.V.~Makarov, N.A.~Asriyan,
A.A.~Elistratov,
Phase alignment in a lattice of exciton-polaritonic Bose-Einstein condensates,
Phys. Lett. A 568 (2026) 131221.
\url{https://doi.org/10.1016/j.physleta.2025.131221}.
%
\bibitem{Quantum}
N.A.~Asriyan, A.A.~Elistratov, Yu.E.~Lozovik,
Mean field study of 2D quasiparticle condensate formation in presence of strong decay,
Quantum 7 (2023) 1144 (2023). 
\url{https://doi.org/10.22331/q-2023-10-16-1144}.
%
\bibitem{Utesov}
O. I. Utesov, M. Park, D. Choi, S. Choi, S. I. Park, S. Kang, J. D. Song, A. N. Osipov, A. V. Yulin, Y.-H. Cho, H. Choi, I. S. Aranson, and 
S. V. Koniakhin,
Communications Physics 8 (2025) 286 (2025).
\url{https://doi.org/10.1038/s42005-025-02198-8}.
%
\bibitem{Xiantao}
X. Li,
Markovian embedding procedures for non-Markovian stochastic Schrödinger equations,
Phys. Lett. A 387 (2021) 127036.
\url{https://doi.org/10.1016/j.physleta.2020.127036}.
%
\bibitem{JRLR2022}
A.D.~Alliluev, D.V.~Makarov,
Dynamics of a nonlinear quantum oscillator under non-Markovian pumping,
J.~Russ.~Laser Res. 43 (2022) 71. 
\url{https://doi.org/10.1007/s10946-022-10024-7}.
%
\bibitem{Sunflower}
G. Christmann, G. Tosi, N. G. Berloff, P. Tsotsis, P. S. Eldridge, Z.~Hatzopoulos, P.~G. Savvidis, and J. J. Baumberg,
Polariton ring condensates and sunflower ripples in an expanding quantum liquid, Phys. Rev. B 85 (2012) 235303.
\url{https://doi.org/10.1103/PhysRevB.85.235303}
%
\bibitem{Fontaine}
Q.~Fontaine, D.~Squizzato, F.~Baboux, I.~Amelio, A.~Lemaitre, M.~Morassi, I.~Sagnes, L.L.~Gratiet, A.~Harouri, M.~Wouters, I.~Carusotto, A.~Amo, M.~Richard., A. Minguzzi, L.~Canet, S.~Ravets, J.~Bloch,
Kardar–Parisi–Zhang universality in a one-dimensional polariton condensate,
Nature 608 (2022) 687.
\url{https://doi.org/10.1038/s41586-022-05001-8}.


\end{thebibliography}
\end{document}